\documentclass[aps,prb,twocolumn,showpacs,groupedaddress]{revtex4-1}
\usepackage{graphicx}  
\usepackage{dcolumn}   
\usepackage{bm}        
\usepackage{amssymb}   
\usepackage{amsmath}
\usepackage{braket}
\usepackage[mathscr]{euscript}
\usepackage{color}
\begin{document}
\title{Magnetic adatoms in two and four terminal graphene nanoribbons: A comparison between their spin polarized transport}
\author{Sudin Ganguly}
\email{sudin@iitg.ernet.in}
\author{Saurabh Basu}
\email{saurabh@iitg.ernet.in}
\affiliation{Department of Physics, Indian Institute of Technology Guwahati\\ Guwahati-781039, Assam, India}
\begin{abstract}
We study
the charge and spin transport in two and four terminal graphene nanoribbons (GNR) decorated with random distribution of magnetic adatoms. The inclusion of the magnetic adatoms generates only the $z$-component of the spin polarized conductance via an exchange bias in the absence of Rashba spin-orbit interaction (SOI), while in presence of Rashba SOI, one is able to create all the three ($x$, $y$ and $z$) components. This has important consequences for possible spintronic applications. The charge conductance shows interesting behaviour near the zero of the Fermi energy. Where in presence of magnetic adatoms the familiar plateau at $2e^2/h$ vanishes, thereby transforming a quantum spin Hall insulating phase to an ordinary insulator. The local charge current and the local spin current provide an intuitive idea on the conductance features of the system. We found that, the local charge current is independent of Rashba SOI, while the three components of the local spin currents are sensitive to Rashba SOI. Moreover the fluctuations of the spin polarized conductance are found to be useful quantities as they show specific trends, that is, they enhance with increasing adatom densities. A two terminal GNR device seems to be better suited for possible spintronic applications.
\end{abstract}
\maketitle
\section{\label{sec1}Introduction}
Graphene-based nanostructures have attracted a wide attention owing to their several interesting electronic and transport properties \cite{novo,neto,zhang,vp,jun,lin,du} for a decade. Unconventional quantum Hall effect \cite{novo,zhang,vp}, half metallicity \cite {jun,lin}, high carrier mobility \cite{du}, such interesting features make graphene as promising candidates for nanoelectronics and spintronics applications. The recent fabrication of freestanding graphene nanoribbons (GNRs) \cite{meyer,moro} has generated  renewed  interest in carbon-based materials with exotic properties. GNRs are basically a single strips of graphene. The electronic properties \cite{fujita,saito} of graphene nanoribbons depend on the geometry of the edges and lateral width of the nanoribbons \cite{nakada}. According to the edge termination type, mainly there are two kinds of GNR, namely armchair graphene nanoribbon (AGNR) and zigzag graphene nanoribbon (ZGNR).

Since the edges play an important role in determining the electronic properties of GNR, they offer a variety of possibilities for tunable electronic properties, such as edge modulation by inorganic atoms, molecules or radicals \cite{wang,hzheng,song,hzeng}, application of transverse electric fields \cite{xh}, adsorption or doping of atoms or molecules \cite{hz,longo,chan,av,dw,kan,sevin,rigo,brito} etc. Metal atoms adsorbed onto graphene sheets also represent a new way for the development of new electronic or spintronic devices. The electronic, structural, and magnetic properties of transition metals (TM) on graphene sheets \cite{chan,av,dw} and graphene nanoribbons (GNR) \cite{kan,sevin,rigo,brito} have been studied extensively, which are mostly based on ab-initio density-functional theory (DFT). The spin dependent transport in GNRs in presence of Rashba SOC has been investigated in some cases, such as spin filtering effect in zigzag GNR \cite{liu}, possible spin polarization directions for GNR with Rashba SOC \cite{chico}, effects of spatial symmetry of GNR on spin polarized transport \cite{qzhang} etc.

Among the metal adatoms, the study of GNRs in presence of transition metals warrants some special attention since TM serve as important catalysts for the synthesis of graphite, CNT, GNR etc. Since TM catalysts (particularly iron) is a common impurity in the graphite \cite{pe}, graphene layers fabricated from graphite are likely to have these impurities. Longo et al. have showed that the behaviour of Fe atom in a GNR is magnetic, in contrast to the behavior found in graphene \cite{av,ejg}. Mao et al. \cite{mao} showed that adsorption of Fe on graphene makes graphene metallic and generates 100\% spin polarization. Basically the study of TM adsorption on graphene \cite{cao,nk}
shows possible applications in the realization of graphene-based electronic and spintronic devices. Motivated by the above studies, it will be highly desirable to explore the spintronic behaviour of TM  adsorption on graphene and graphene-based structures.

In the present work, we explore the charge and spin transport properties of ZGNR decorated by magnetic adatoms.  For this, we consider a case where TM (particularly Fe) are adsorbed onto ZGNR. The resulting broken structural symmetry gives rise to a Rashba spin-orbit interaction (RSOI) and the hybridization between the carbon $\pi$ state and the 3$d$-shell states of magnetic adatoms generates a macroscopic exchange field \cite{jiang}. A hall conductivity, $\sigma^H_{xy}$ of magnitude $2e^2/h$ is observed for the case where the Fermi energy lies in the bulk gap. First principle calculations report a bulk gap of almost 5.5 meV in Fe adsorbed GNRs, which should be possible to verify in experiments \cite{jiang}. Further, in order to avoid any complicacy, such as adatom-adatom interaction, spin-spin correlation etc., we consider very small concentration of magnetic adatoms, so that any interactions other than RSOI and exchange field will not be present.

We organize our paper as follows. In the following section, we present, for completeness, the theoretical formalism leading to the expressions for the  two terminal charge and spin polarized conductances and four terminal longitudinal and spin Hall conductances using the well known Landauer-B\"{u}ttiker formula. After that we include an elaborate discussion of the results. Here, we have tried to resolve few queries, how the spin polarized conductance behaves in the two terminal case, whether there is any similarity between the two terminal spin polarized conductance and four terminal spin Hall conductance. We have also included an interesting comparison for the charge conductance properties for the case of two and four terminal ZGNR.

\section{\label{sec2}Theoretical formulation and model}

\begin{figure}[h]
\begin{center}
\includegraphics[width=0.3\textwidth]{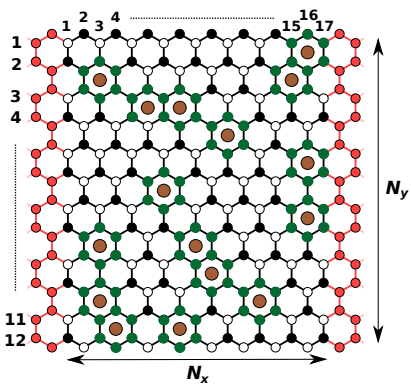}
\caption{(Color online) Schematic view of a two terminal graphene nanoribbon. The black and white circles represent the A and B sublattices of graphene. The brown circles are the magnetic adatoms. The green circles are the affected site due to magnetic adatoms. The black lines surrounding the magnetic atoms correspond to nearest neighbour hopping and Rashba SOC. Rest of the black lines contain only nearest neighbour hopping. The leads are attached at both ends, which are denoted by red color and are semi-infinite in nature. The leads are free of any kind of SOC. $N_x$ and $N_y$ are the length and width of the nanoribbon respectively.}
\label{setup}
\end{center}
\end{figure}

To begin with we describe the geometry of the system and make our notations clear. We consider a graphene sheet adsorbed with magnetic atoms, which induces Rashba SOI and generates an exchange field. The effective tight-binding Hamiltonian for graphene with such adatoms is given by \cite{jiang},

\begin{eqnarray}
H&=& - 
t\sum\limits_{\langle ij\rangle}c_i^{\dagger} c_j + 
i\lambda_R\sum\limits_{\langle ij\rangle\in\mathcal{R}}c_i^{\dagger} \left( {\vec{\sigma}} \times {\bf\hat{d}}_{ij}\right)_z c_j \nonumber \\ &+& 
\lambda_{EB}\sum\limits_{i\in \mathcal{R}} c_i^{\dagger}\sigma_zc_i
\label{h2}
\end{eqnarray} 
where $c_i^{\dagger}=\left(c_{i\uparrow}^{\dagger} \quad c_{i\downarrow}^{\dagger}\right)$ is the creation operator of electrons at site $i$. The first term is the nearest neighbour hopping term, with a hopping strength, $t=2.7$ eV. The second term is the nearest neighbour Rashba term which explicitly violates $z\rightarrow -z$ symmetry. This term is induced by the adatoms residing on the set of hexagons $\mathcal{R}$ that are inhabited by the magnetic adatoms. The third term is the exchange bias that (as shown in Fig.\ref{setup}) originates due to magnetic adatoms.

\subsection{Two terminal (2T) GNR: formulation of charge and spin polarized conductances}
The zero temperature conductance, $G$ which
denotes the charge transport measurements, is related with
the transmission coefficient as in \cite {land_cond,land_cond2},
\begin{equation}
G = \frac{e^2}{h} T(E)
\end{equation}
The Transmission coefficient can be calculated via \cite{caroli,Fisher-Lee},

\begin{equation}
T = \text{Tr}\left[\Gamma_R {\cal G}_R
  \Gamma_L {\cal G}_A\right]
\end{equation}
${\cal G}_{R(A)}$ is the retarded (advance) Green's function. $\Gamma_{L(R)}$ are the coupling matrices representing the
coupling between the central region and the left (right) lead. They are
defined by the relation \cite{dutta},
\begin{equation}
\Gamma_{L(R)} = i\left[\Sigma_{L(R)} -
  (\Sigma_{L(R)})^\dagger\right]
\end{equation}
Here $\Sigma_{L(R)}$ is the retarded self-energy associated with the left (right) lead. The self-energy contribution is computed
by modeling each terminal as a semi-infinite perfect wire \cite{nico}.

We define the spin polarized conductance as,
\begin{equation}
G^s_\alpha = \frac{\hbar}{2e}\left[\frac{I_R^\alpha}{V_L-V_R}\right]
\end{equation}
$I_R^\alpha \;(\alpha=x,y,z)$ is the spin current flowing through right lead and $V_{L/R}$ is the potential at the left/right lead.

The spin polarized current can be calculated using \cite{chang,roche},
\begin{equation}
I_R^\alpha = \frac{e^2}{h} \text{Tr}\left[\hat{\sigma}_\alpha\Gamma_R G_R
  \Gamma_L G_A\right](V_R - V_L)
\end{equation}
where, $\alpha=x,y,z$ and $\sigma$ denote the Pauli matrices.

Fig.\ref{setup} shows the geometry used for the calculations of charge and spin polarized conductances. Fig.\ref{setup} is the setup corresponds to ZGNR. The length and width of these systems can be determined as shown in the given figure. In Fig.\ref{setup}, the width is, $N_y = 12$ and the length is, $N_x = 21$. Thus we can denote the zigzag setup by $N_x$Z-$N_y$A $=$ 17Z-12A. This nomenclature for denoting the dimensions of the nanoribbon will be followed throughout the paper.

The black and white circles stand for the A and B sublattices of graphene. The brown circles are the magnetic adatoms. The green circles are the affected site due to magnetic adatoms. The black lines surrounding the adatoms correspond to the nearest neighbour hopping and the Rashba SOC. Rest of the black lines denote only nearest neighbour hopping. The leads are semi-infinite in nature, attached at both ends and are denoted by red color. The leads are considered to describes by a pure tight binding graphene lattice and hence are free of any kind of SOC.

\subsection{Four terminal (4T) GNR: formulation of longitudinal and spin Hall conductances}

\begin{figure}[h]
\begin{center}
\includegraphics[width=0.3\textwidth]{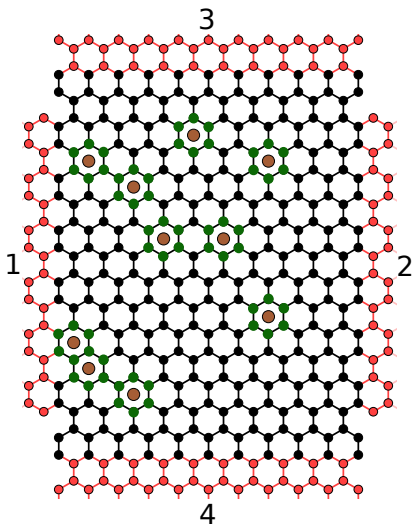}
\caption{(Color online) Schematic view of a four terminal graphene nanoribbon. The black and white circles represent the A and B sublattices of graphene. The brown circles are the magnetic adatoms. The green circles are the affected sites due to magnetic adatoms. The black lines surrounding the Au atoms correspond to nearest neighbour hopping and Rashba SOC. Rest of the black lines contain only nearest neighbour hopping. The leads are attached at the four sides, which are denoted by red color and are semi-infinite in nature. The leads are free of any kind of SOC.}
\label{setup4}
\end{center}
\end{figure}

In order to observe the spin Hall conductance, a charge current is allowed to flow between terminal 1 and 2, and a spin current is observed to flow along the transferred direction of the rectangular sample, that is between terminals 3 and 4 as shown in Fig.\ref{setup4}.

In case of four terminal device, the longitudinal and spin Hall conductances are defined as follows,
\begin{equation}
G = \frac{I^q_2}{V_1-V_2}\quad G_{SH}^\alpha = \frac{\hbar}{2e}\left[\frac{I_3^\alpha}{V_1-V_2}\right]
\end{equation}
where, $I^q_2$ is the charge current flowing through terminal 2 and $I_3^\alpha \;(\alpha=x,y,z)$ is the spin current polarized in a particular direction $\alpha$ and flowing through terminal 3. $V_i$ is the potential at the $i$-th lead.

Following the Landauer-B\"{u}ttiker formula \cite{land_cond,land_cond2}, the charge and spin currents can be calculated from the following expression \cite{chang,roche},
\begin{equation}
I^{\alpha}_p = \frac{e^2}{h} \sum\limits_q \text{Tr}\left[\hat{\sigma}_\alpha\Gamma_q {\cal G}_R
  \Gamma_p {\cal G}_A\right](V_p - V_q)
\label{current}
\end{equation}
where, $\hat{\sigma}_\alpha=(\sigma_0,\sigma_x,\sigma_y,\sigma_z)$. $\sigma_0$ is a $2\times2$ identity matrix and $\sigma_x,\sigma_y,\sigma_z$ are the Pauli matrices. $\sigma_0$ in Eq.\ref{current} gives the usual charge current, while the Pauli matrices yield the spin currents polarized in different directions ($x$, $y$ and $z$).

Since leads 3 and 4 are voltage probes, $I^q_3=I^q_4=0$. On the other
hand, as the currents in various leads depend only on voltage
differences among them, we can set one of the voltages to zero without
any loss of generality. Here we set $V_2 = 0$.

\section{\label{sec3}Results and Discussions}
We have studied the charge and spin conductance properties in ZGNR decorated by magnetic adatoms and compared the results for two and four terminal devices.

Before embarking on the results, we briefly describe the values of different parameters used in our calculation. We set the hopping term, $t=2.7$ eV. Throughout our work, we take the ZGNR setup as 89Z-48A for the two terminal case. While for the four terminal device it is 85Z-52A. All the energies are measured in unit of $t$. The charge conductance is measured in units of $\frac{e^2}{\hbar}$ and the spin polarized conductance is measured in units of  $\frac{e}{4\pi}$ for the 2T case. The longitudinal and spin Hall conductances are measured in units of $\frac{e^2}{\hbar}$ and $\frac{e}{4\pi}$ respectively for the 4T case. Also the lattice constant, $a$ is taken to be unity. All the measurable quantities are averaged over 50 independent random-adatom configurations for different adatom concentrations, $n_{ad}$. In this work, we have considered three different adatom concentrations, namely, $n_{ad} = 0.025,0.05$ and $0.1$. We have checked that 50 configurations are adequate in the present context, especially, since the adatom densities considered here are small. For most of our numerical calculations we have used KWANT \cite{kwant}.

\subsection{Two terminal}
\begin{figure}[h]
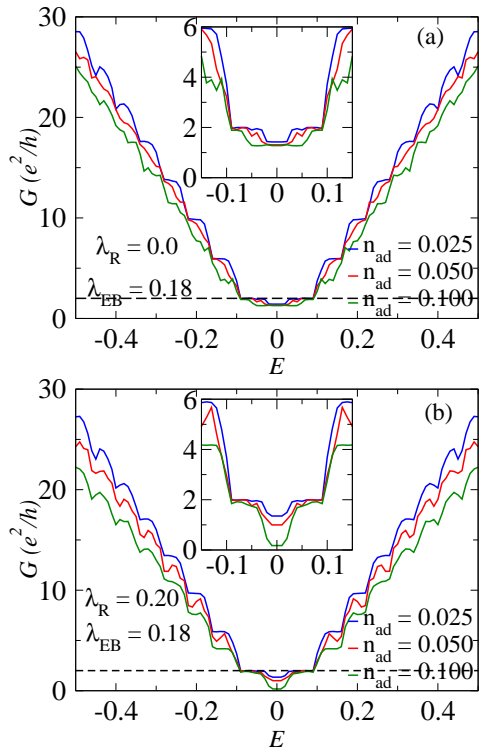

\begin{center}
\includegraphics[width=0.35\textwidth]{fig3a.eps}\quad\includegraphics[width=0.35\textwidth]{fig3b.eps}
\caption{(Color online) The two terminal charge conductance, $G$ is plotted as a function of the Fermi energy, E (a) in the absence of Rashba SOI and (b) in the presence of Rashba SOI with strength, $\lambda_R = 0.2$. The strength of the exchange field is fixed in both the cases at $\lambda_{EB} =0.18$. To visualize the behaviour of $G$ near the zero of the Fermi energy, we plot the variation of $G$ for a small range of the Fermi energy as shown in the insets.}
\label{2t_cond}
\end{center}
\end{figure}

The behaviour of charge conductance, $G$ for two terminal case is studied as a function of the Fermi energy, $E$ as shown in Fig.\ref{2t_cond} for two different cases, $\lambda_R=0$ in Fig.\ref{2t_cond}(a) and $\lambda_R=0.2$ in Fig.\ref{2t_cond}(b). The strength of the exchange field is kept fixed at $\lambda_{EB}=0.18$. Corresponding to these parameters, the insulating bulk states are clearly discernible from the chiral conducting edge states \cite{jiang}. The variation of $G$ as a function of the Fermi energy is quite similar for two different strengths of the Rashba SOI. $G$ is symmetric around the zero of the Fermi energy and shows a nice necklace-type pattern for both the case. However, the behaviour of $G$ close to zero of the Fermi energy is different for the two different values of $\lambda_R$. In order to visualize the behaviour of $G$ close to zero of the Fermi energy, we plot $G$ for a small range of the Fermi energy as shown in the inset in Fig.\ref{2t_cond}. In the absence of Rashba SOI, $G$ always stays finite. On the other hand, in presence of Rashba SOI, $G$ tends to vanish values in the vicinity of $E=0$ as we increase the adatom concentration. It identically vanishes at $n_{ad} = 0.1$. The $2e^2/h$ plateau in pristine graphene in presence of Rashba SOI is believed to be a signature of the quantum spin Hall insulating phase protected by the time reversal symmetry. In presence of the exchange bias, the time reversal symmetry is explicitly violated leading to the disappearance of the plateau and giving rise to features of an ordinary insulator.

\begin{figure}[h]
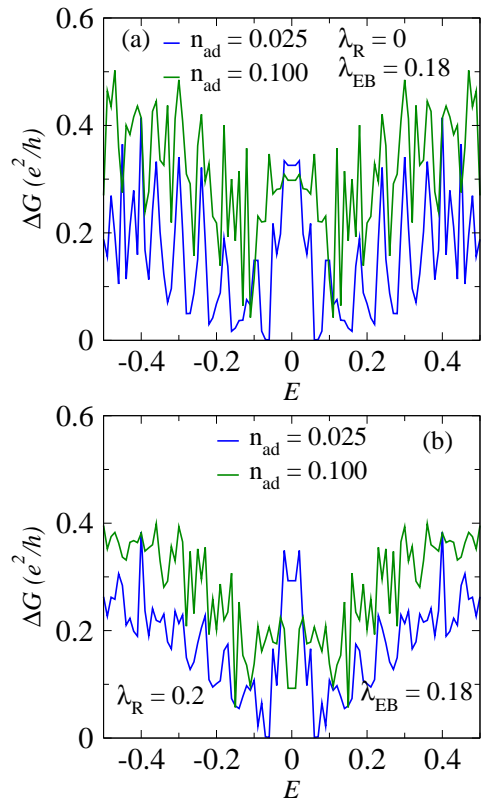

\begin{center}
\includegraphics[width=0.35\textwidth]{fig4a.eps}\\\includegraphics[width=0.35\textwidth]{fig4b.eps}
\caption{(Color online) The fluctuations in the charge conductance, $\Delta G$ is plotted as a function of the Fermi energy, $E$ for (a) $\lambda_R = 0$ and (b) $\lambda_R=0.2$. Only two different adatom concentrations are taken, namely, $n_{ad} = 0.025$ and 0.1 for clarity.}
\label{2t_dcond}
\end{center}
\end{figure}

We did not show the fluctuations in the charge conductance in Fig.\ref{2t_cond} in order to avoid cluttering of data. However, the fluctuations in the charge conductance, $\Delta G$ is plotted as a function of the Fermi energy separately as shown in Fig.\ref{2t_dcond} for clarity. Fig.\ref{2t_dcond}(a) shows the variation of $G$ in the absence of Rashba SOI and Fig.\ref{2t_dcond}(b) for a Rashba strength, $\lambda_R = 0.2$. In both cases, fluctuations increases with increasing the adatom concentration. Rashba SOI suppresses the fluctuations in the charge conductance as seen from Fig.\ref{2t_dcond}(b). The spike-like behaviour in $\Delta G$ is due to the finite number of open channels in the leads and may be due to the spin precision effect \cite{sudin,lsheng}. For clarity, we have taken only two different adatom concentrations, namely $n_{ad}=$ 0.025 (denoted by blue color) and 0.1 (denoted by green color).

It is expected that in the absence of Rashba SOI, the exchange field generates only the $z$-component of the spin polarized conductance, since the exchange field contains only the $z$-component of the Pauli spin matrix (see Eq.\ref{h2}). However, the inclusion of the Rashba SOI generates all the three components of the spin polarized components.

\begin{figure}[h]
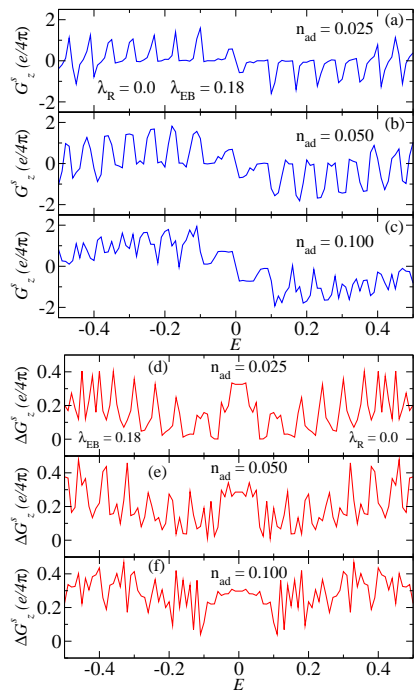

\begin{center}
\includegraphics[width=0.3\textwidth]{fig5a.eps}\\\includegraphics[width=0.3\textwidth]{fig5b.eps}
\caption{(Color online) (a-c)The two terminal spin polarized conductance, $G^s_z$ is plotted as a function of the Fermi energy in the absence of Rashba SOI for three different adatom concentrations, namely $n_{ad} =$0.025, 0.05 and 0.1. (d-f) Their corresponding fluctuations are plotted as a function of $E$.}
\label{2t_gsz}
\end{center}
\end{figure}

Fig.\ref{2t_gsz}(a-c) shows the variation of the $z$-component of the spin polarized component, $G^s_z$ (in units of $e/4\pi$) as a function of the Fermi energy, $E$ in the absence of Rashba SOI for three different adatom concentrations. $G^s_z$ is antisymmetric as a function of the Fermi energy owing to the electron-hole symmetry \cite{chico,moca}. Though we consider very dilute adatom concentrations, $G^s_z$ acquires quite a large value in the absence of Rashba SOI. The spike-like nature as discussed before is due to the finite number of open channels in the leads. The fluctuations in $G^s_z$ denoted by $\Delta G^s_z$ is plotted as a function of the Fermi energy, $E$ as shown in Fig.\ref{2t_gsz}(d-f). $\Delta G^s_z$ is symmetric about $E=0$. However, in the absence of Rashba SOI, $\Delta G^s_z$ is more or less independent of the adatom concentration.

\begin{figure}[h]
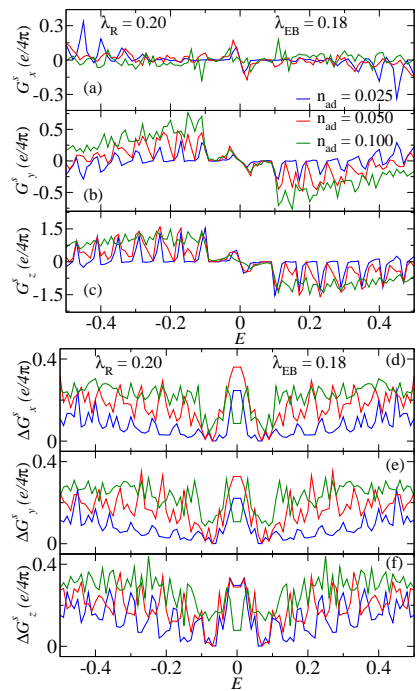

\begin{center}
\includegraphics[width=0.3\textwidth]{fig6a.eps}\\\includegraphics[width=0.3\textwidth]{fig6b.eps}
\caption{(Color online) In presence of Rashba SOI, the three components of the spin polarized conductance (a) $G^s_x$, (b) $G^s_y$ and $G^s_z$ are plotted as a function of the Fermi energy. (d-f) Their fluctuations are plotted as a function of $E$.  }
\label{2t_gsz_r}
\end{center}
\end{figure}
The behaviour of all the three components of spin polarized conductance as a function of $E$ in presence of Rashba SOI is plotted in Fig.\ref{2t_gsz_r}(a-c). The magnitude of the spin polarized conductance increases with increasing the adatom concentration. The $z$-component, namely $G^s_z$ has a higher magnitude than the other two components of the spin polarized conductance. However, all of them are antisymmetric about $E=0$ as they should be. The magnitude of $G^s_z$ in presence of Rashba SOI is lower than the values corresponding to that of in the absence of Rashba SOI. The spike-like feature is mostly prominent in the behaviour of $G^s_z$ and a little in $G^s_y$. The behaviour of the corresponding fluctuations in the spin polarized conductance is shown in Fig.\ref{2t_gsz_r}(d-f). As the adatom concentration is increased, the fluctuations also increase. But the most interesting feature about the fluctuations is that all the three components, that is $\Delta G^s_i (i=x,y,z)$ have a unique behaviour as elaborated in the following. Even if the different components of the spin polarized conductance have different variations as a function of $E$ and have different orders of magnitude for a particular strength of the exchange field in presence of Rashba SOI, yet the orders of magnitude of their fluctuations are almost same. Further, as usual the fluctuations increases with increasing adatom concentration.

The inclusion of magnetic adatoms generates the exchange field and as a result, the time reversal symmetry is broken. Hence the system will no longer have a Kramer's doublet. This phenomena can be justified through the density of states (DOS) for different species, that is DOS for up and down spins. We denote the total density of states by DOS, the density of states coming from spin up electron by UDOS and that from spin down electron by DDOS. We also define the difference between the UDOS and DDOS by DiffDOS. 
\begin{figure}[h]
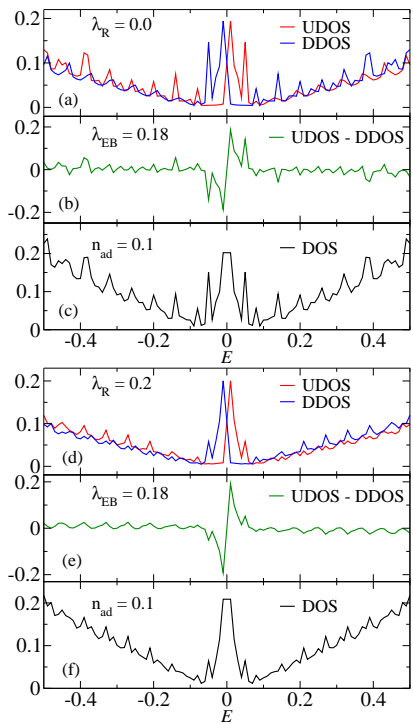

\begin{center}
\includegraphics[width=0.3\textwidth]{fig7a.eps}\\\includegraphics[width=0.3\textwidth]{fig7b.eps}
\caption{(Color online) The density of states (DOS) for the spin up and spin down electrons are plotted as a function of the Fermi energy, $E$ (a) for $\lambda_R = 0$ and (d) for $\lambda_R = 0.2$. The difference between DOS$\uparrow$ and DOS$\downarrow$ are plotted as a function of $E$ (b) for $\lambda_R = 0$ and (e) for $\lambda_R = 0.2$. The behaviour of total DOS is plotted as a function of the Fermi energy (c) for $\lambda_R = 0$ and (e) for $\lambda_R = 0.2$. for clarity we consider only a single adatom concentration, $n_{ad} = 0.1$.}
\label{2t_dos}
\end{center}
\end{figure}

The variation of the DOS as a function of the Fermi energy is hence plotted in Fig.\ref{2t_dos} for a particular strength of the exchange field, $\lambda_{EB}=0.18$. Fig.\ref{2t_dos}(a) shows the behaviour of UDOS and DDOS in the absence of Rashba SOI and Fig.\ref{2t_dos}(d) in presence of Rashba SOI. It is observed that UDOS is antisymmetric with respect to DDOS as a function of the Fermi energy. The difference between UDOS and DDOS, that is DiffDOS is plotted in Fig.\ref{2t_dos}(b) and Fig.\ref{2t_dos}(e) in the absence and presence of Rashba SOI respectively. In both cases UDOS $-$ DDOS is antisymmetric about $E=0$. The total DOS, that is the sum of DOS due to spin up and spin down electrons is plotted in Fig.\ref{2t_dos}(c) and Fig.\ref{2t_dos}(f). DOS is symmetric about $E=0$. 

We can summarize the observations noted from the DOS data, as shown by the following set of properties,
\begin{eqnarray}
\text{UDOS}(E) &=& \text{DDOS}(-E) \nonumber\\
\text{DiffDOS}(E) &=& -\text{DiffDOS}(-E) \\
\text{DOS}(E) &=& \text{DOS}(-E) \nonumber
\label{dos}
\end{eqnarray}

In order to have a deeper look, we have also shown the local charge and spin currents for a fixed Fermi energy. Since the left lead acts as an input to the system, we are interested in the local charge and spin currents which have originated due to the left lead. 

\begin{figure}[h]
\begin{center}
\includegraphics[width=0.23\textwidth]{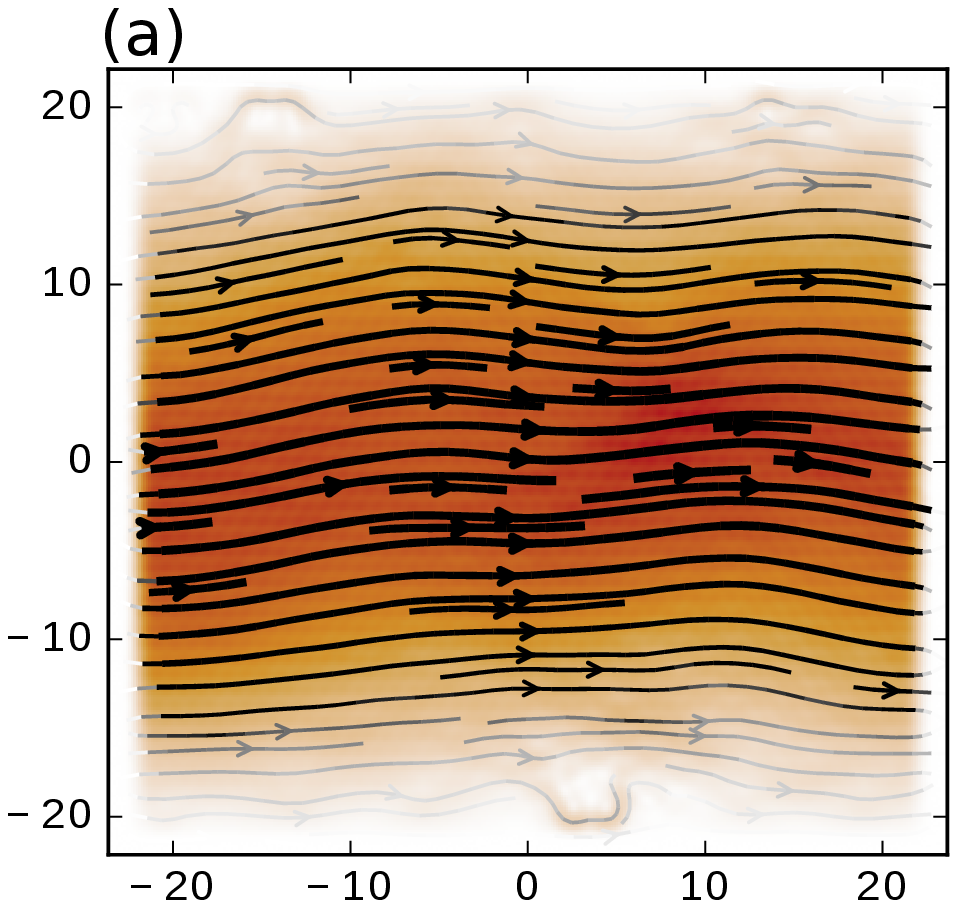}\includegraphics[width=0.23\textwidth]{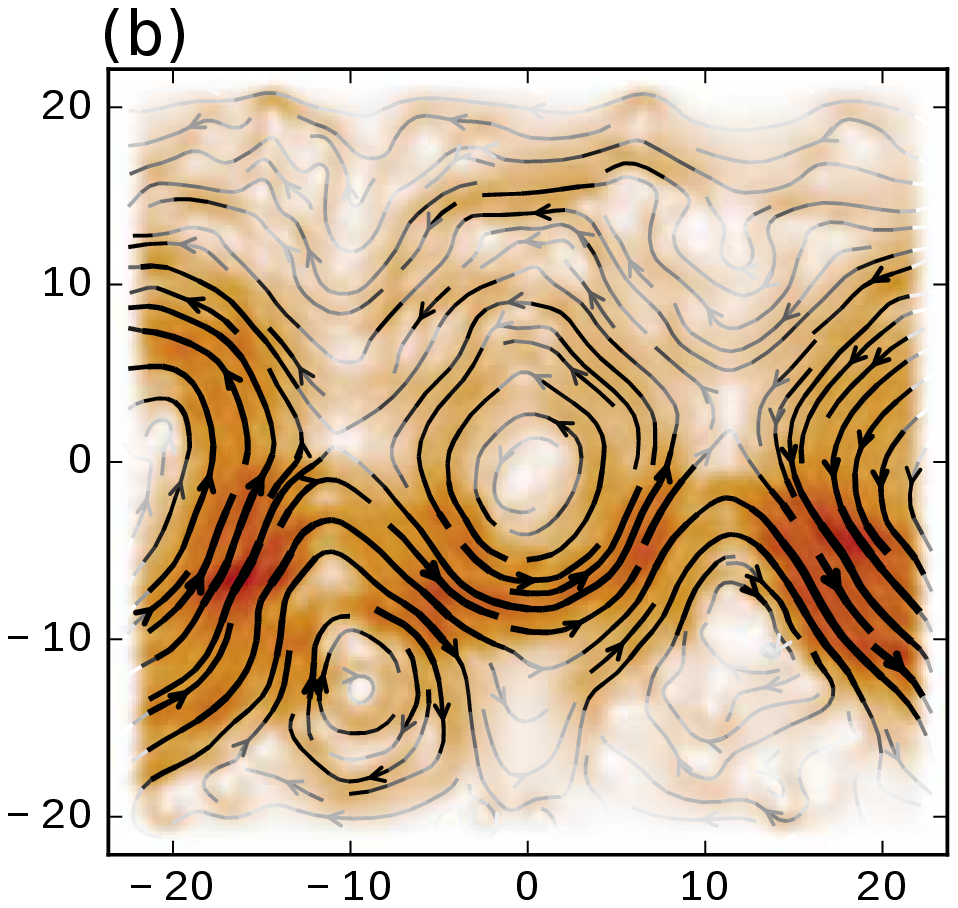}
\caption{(Color online) (a) The local charge current, $J_0$ and (b) the $z$-component of the local spin current are shown in the absence of Rashba SOI. Both the figures are obtained for the adatom concentration $n_{ad} = 0.1$ and we set the Fermi energy at $E = -0.09$. Figure (b) shows presence of some centers around which $J_z$ whirls. }
\label{2t_current}
\end{center}
\end{figure}
Fig.\ref{2t_current}(a) shows the nature of the local charge current, $J_0$ in the absence of Rashba SOI for a fixed value of the Fermi energy, namely $E=-0.09$. Clearly the local charge current is flowing between the left to the right lead without any distortion. In case of spin currents, there are three components, namely $J_x,J_y,J_z$. Since  in the absence of Rashba SOI only $z$-component of the local spin current exits, we have calculated $J_z$ as shown in Fig.\ref{2t_current}(b). The number of paths between left and right leads are less compared to the local charge current. Few of the paths are even circling around certain points signaling a vortex like behaviour.

\begin{figure}[h]
\begin{center}
\includegraphics[width=0.2\textwidth]{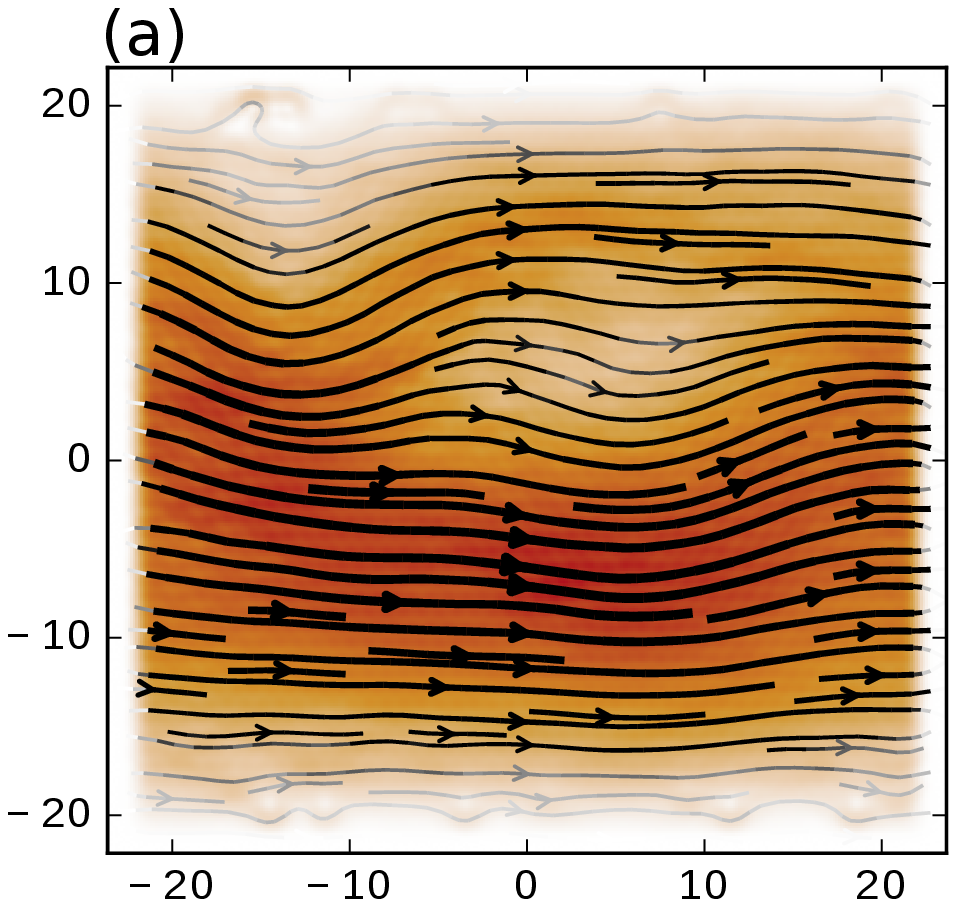}\quad\includegraphics[width=0.2\textwidth]{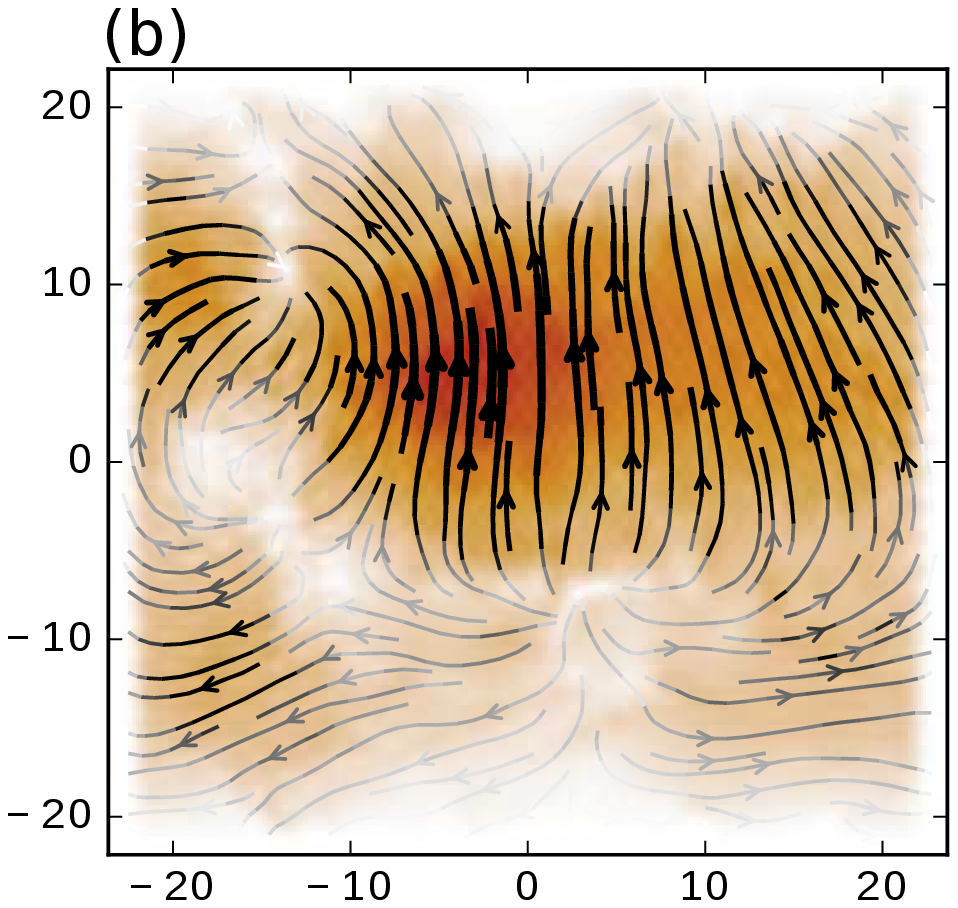}
\includegraphics[width=0.2\textwidth]{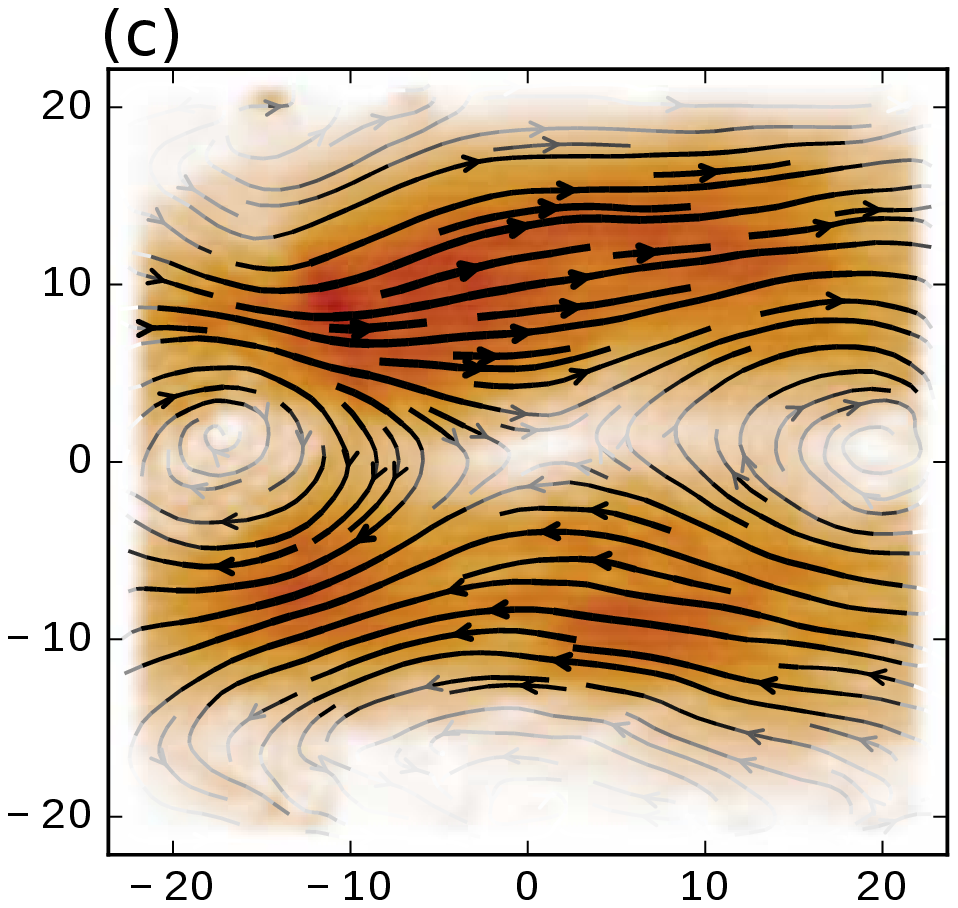}\quad\includegraphics[width=0.2\textwidth]{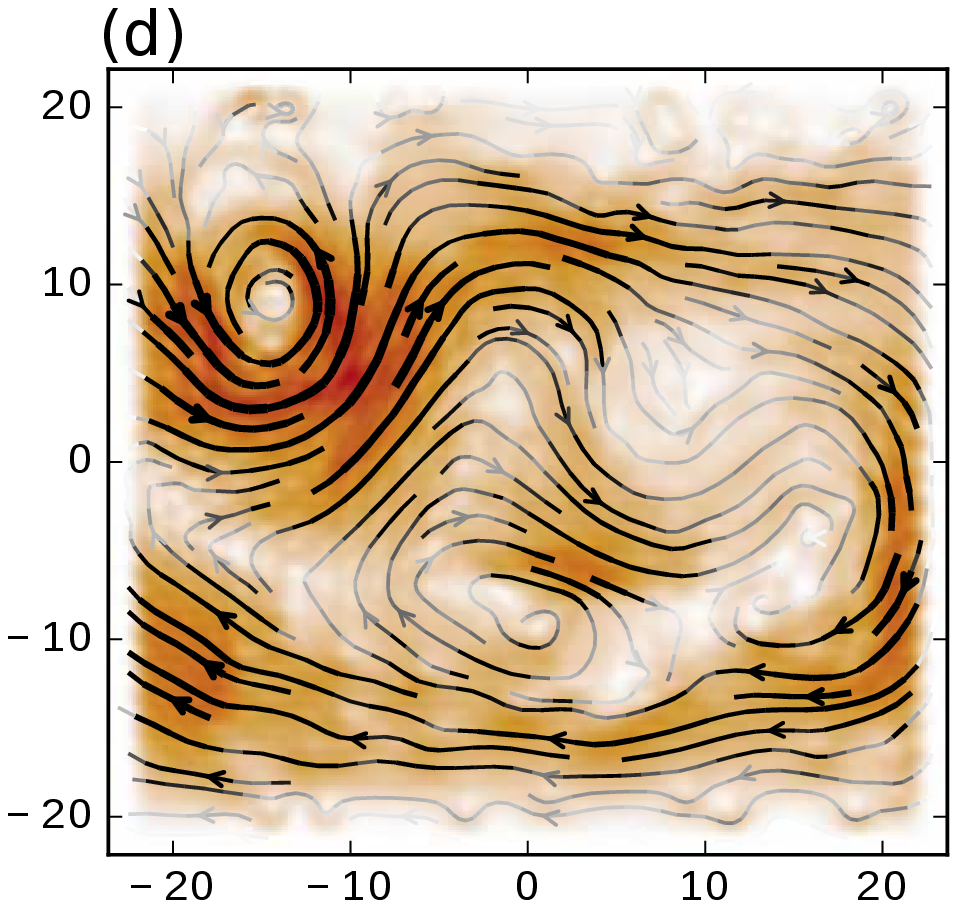}
\caption{(Color online) (a) The local charge current $J_0$, (b) the $x$-component of the local spin current, $J_x$, (c) the $y$-component of the local spin current, $J_y$ and (d) the $z$-component of the local spin current, $J_z$ are shown in the presence of Rashba SOI with strength $\lambda_R = 0.2$. The strength of the exchange field is $\lambda_{EB} = 0.18$. We set the Fermi energy at $E = -0.09$ and the adatom concentration in the present case is $n_{ad} = 0.1$. Centers similar to Fig.\ref{2t_current}(b) have been noticed in the $J_x$, $J_y$ and $J_z$ plots.}
\label{2t_current_lr}
\end{center}
\end{figure}

Since the inclusion of Rashba SOI generates the two other components of the spin polarized conductance, namely the $x$ and $y$ components, it will be meaningful to study the local spin currents of these components. Existence of all of these components should be beneficial for spintronic applications of magnetic adatom decorated GNRs.

For this, we set, as before, the strength of the  Rashba SOI at $\lambda_R=0.2$ and fixed the energy at $E=-0.09$. The local charge and spin currents are shown in Fig.\ref{2t_current_lr}. The local charge current, $J_0$ is again flowing between left and right leads as shown in Fig.\ref{2t_current_lr}(a), which reveals that $J_0$ is almost independent of the Rashba SOI which is understandable. Fig.\ref{2t_current_lr}(b),(c) and (d) show respectively $J_x$, $J_y$ and $J_z$. If we recall Fig.\ref{2t_gsz_r}, where $G^s_i (i=x,y,z)$ are plotted as a function of the Fermi energy, the $x$-component of the spin polarized conductance has the lower magnitude compared to the other two components. This can be explained from Fig.\ref{2t_current_lr}(b). Here we see that the number of clear paths between left and right leads are very less compared to the paths corresponding to $J_y$ and $J_z$. As a result, $G^s_x$ has lesser magnitude than $G^s_y$ and $G^s_z$.

One point should be mentioned here that all the plots in Fig.\ref{2t_current} and Fig.\ref{2t_current_lr} are obtained for a single configuration corresponding to $n_{ad}=0.1$. Another configuration or taking an average over several configurations may change the plots, but the qualitative feature will remain same.

\subsection{Four terminal}
Having emphasized upon the conductance characteristics of a 2T GNR, it is useful to compare and contrast with respect to the 4T devices.

To begin with the results in case of four terminal device, we shall remind ourselves that the setup for measuring the longitudinal conductance $(G)$ and spin Hall conductance $(G_{SH})$. As shown in Fig.\ref{setup4}, an electric current is allowed to pass between terminal 1 and 2, and the longitudinal conductance is measured. Terminals 3 and 4 are the voltage probes, hence there will be no charge current flowing through them. The spin Hall conductance is measured between terminals 3 and 4.

\begin{figure}[h]
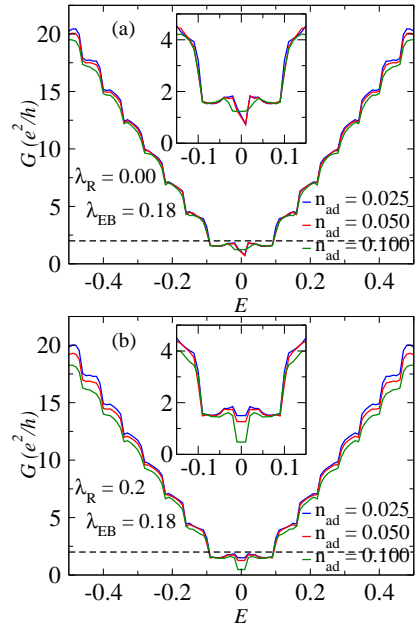

\begin{center}
\includegraphics[width=0.3\textwidth]{fig10a.eps}\\\includegraphics[width=0.3\textwidth]{fig10b.eps}
\caption{(Color online) The four terminal charge conductance, $G$ is plotted as a function of the Fermi energy, $E$ (a) in the absence of Rashba SOI and (b) in the presence of Rashba SOI with strength, $\lambda_R = 0.2$. The strength of the exchange field is fixed in both the cases at $\lambda_{EB} =0.18$. To visualize the behaviour of $G$ near the zero of the Fermi energy, we plot the variation of $G$ for a small range of the Fermi energy as shown in the inset.}
\label{4t_cond}
\end{center}
\end{figure}
The behaviour of the charge conductance, $G$ for a 4T case is studied as a function of the Fermi energy ($E$) as shown in Fig.\ref{4t_cond} for two different cases as earlier, namely, $\lambda_R=0$ in Fig.\ref{4t_cond}(a) and $\lambda_R=0.2$ in Fig.\ref{4t_cond}(b). The strength of the exchange field is kept fixed as in case of two terminal case, that is at $\lambda_{EB}=0.18$. $G$ is symmetric around the zero of the Fermi energy and shows a nice necklace-type pattern for both the case. However, the behaviour of $G$ close to zero of the Fermi energy is different for two different values of $\lambda_R$. We plot $G$ for a small range of the Fermi energy in the vicinity of $E=0$ as shown in the inset in Fig.\ref{4t_cond}. In the absence of Rashba SOI, $G$ always remain non-zero. On the other hand, in presence of Rashba SOI, $G$ tends to have lower values near $E=0$ as we increase the adatom concentration. One can expect an insulating behaviour about $E=0$ in presence of Rashba SOI if we increase the adatom concentration beyond $n_{ad} = 0.1$. There is a crucial difference between the 2T and 4T devices with regard to the plateau at $2e^2/h$ in the vicinity of the zero of the Fermi energy implying the existence of a quantum spin Hall phase. As discussed earlier, there is a fairly flat plateau for a 2T GNR without magnetic adatoms, which is visibly absent for a 4T GNR (see Fig.\ref{4t_cond} and the insets). This is another minor, where in a 2T device, we have observed $G$ to completely vanish at $n_{ad}=0.1$, while in the 4T GNR, it would eventually vanish for $n_{ad} > 0.1$.

\begin{figure}[h]
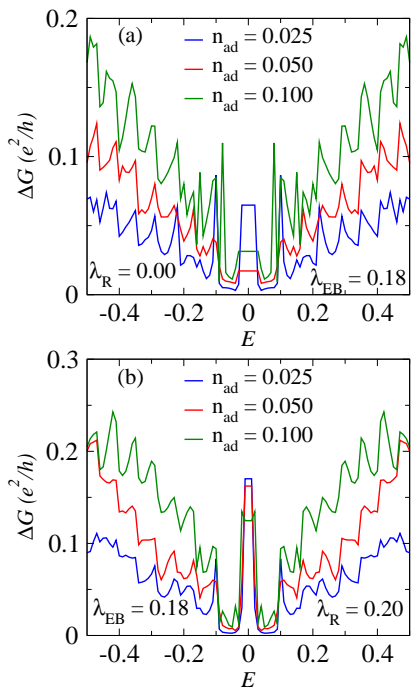

\begin{center}
\includegraphics[width=0.3\textwidth]{fig11a.eps}\\\includegraphics[width=0.3\textwidth]{fig11b.eps}
\caption{(Color online) The fluctuations in the charge conductance, $\Delta G$ is plotted as a function of the Fermi energy, $E$ for (a) $\lambda_R = 0$ and (b) $\lambda_R=0.2$. Three different adatom concentrations are taken, namely, $n_{ad} = 0.025$, 0.05 and 0.1. }
\label{4t_dcond}
\end{center}
\end{figure}

The fluctuations in the charge conductance, $\Delta G$ is plotted in Fig.\ref{4t_dcond}(a) in the absence of Rashba SOI and in Fig.\ref{4t_dcond}(b) in presence of Rashba SOI. In each of the plots, $\Delta G$ falls off as we come close to $E=0$, from either side of the band and again increases near $E=0$. The fluctuations increases as we increase the adatom concentration. By comparison the two plots, we observe that the inclusion of Rashba SOI causes larger fluctuations than the corresponding case where Rashba SOI is absent.

\begin{figure}[h]
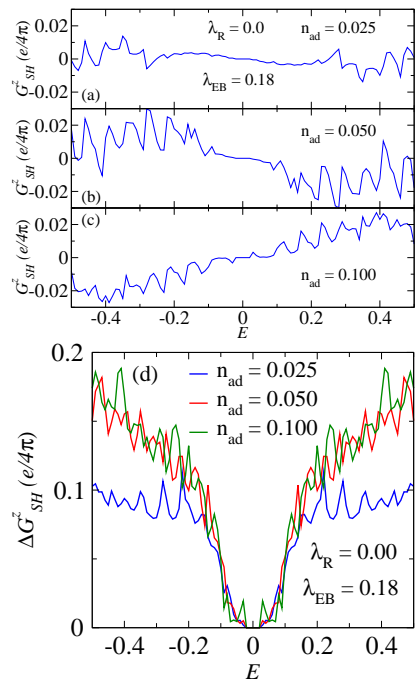

\begin{center}
\includegraphics[width=0.3\textwidth]{fig12a.eps}\\\includegraphics[width=0.3\textwidth]{fig12b.eps}
\caption{(Color online) (a-c)The four terminal spin polarized conductance, $G^s_z$ is plotted as a function of the Fermi energy, $E$ in the absence of Rashba SOI for three different adatom concentrations. (d) Their corresponding fluctuations are plotted as a function of $E$.}
\label{4t_gsz}
\end{center}
\end{figure}

Same as that of the spin polarized conductance, the spin Hall conductance has also three components, namely $G_{SH}^x$, $G_{SH}^y$ and $G_{SH}^z$. Since the $x$ and $y$-component of the spin Hall conductance are absent in the absence of Rashba SOI, we have shown only the variation of $G_{SH}^z$ in units of $e/4\pi$ as a function of the Fermi energy as shown in Fig.\ref{4t_gsz}(a-c)and corresponding fluctuations in Fig.\ref{4t_gsz}(d).

$G_{SH}^z$ is plotted for three different adatom concentrations separately for better clarity. The behaviour of $G_{SH}^z$ is antisymmetric about $E=0$ and the magnitude (irrespective of the sign factor) increases as the adatom density is enhanced. The maximum values of $G_{SH}^z$ are observed on either side of the zero of the Fermi energy and close to $E=0$, the value decreases. The corresponding fluctuations are shown in a single plot in Fig.\ref{4t_gsz}(d). As expected, with increasing the adatom concentration, $\Delta G_{SH}^z$ increases. Another important point to be noted here is that there is a finite region about the zero energy where $G_{SH}^z$  and $\Delta G_{SH}^z$ are strictly zero irrespective of the adatom concentration which is owing to the single channel transmission \cite{chico,liu}.

\begin{figure}[h]
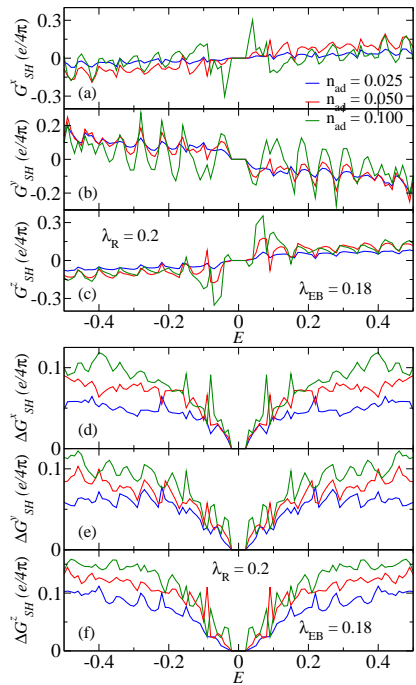

\begin{center}
\includegraphics[width=0.3\textwidth]{fig13a.eps}\\\includegraphics[width=0.3\textwidth]{fig13b.eps}
\caption{(Color online) In presence of Rashba SOI, the three components of the spin polarized conductance (a) $G^s_x$, (b) $G^s_y$ and $G^s_z$ are plotted as a function of the Fermi energy, $E$. (d-f) Their fluctuations are plotted as a function of $E$.  }
\label{4t_gsz_lr}
\end{center}
\end{figure}

The effect of the inclusion of Rashba SOI on the three components of the spin Hall conductance is observed in Fig.\ref{4t_gsz_lr}(a-c). All of them are antisymmetric about $E=0$. Their magnitude increases with increasing the adatom concentration. Though the behaviour of the different components of the spin Hall conductance are different in nature as a function of the Fermi energy, their order of magnitude is similar. The fluctuations in the spin Hall conductance are plotted in Fig.\ref{4t_gsz_lr}(d-f) as a function of the Fermi energy. as observed the case in the absence of Rashba SOI, in the present case, $G_{SH}^x$, $G_{SH}^y$ and $G_{SH}^z$ have similar variations with $E$, that is they increase with increasing adatom concentration. further, their orders of magnitude are also same.

\begin{figure}[h]
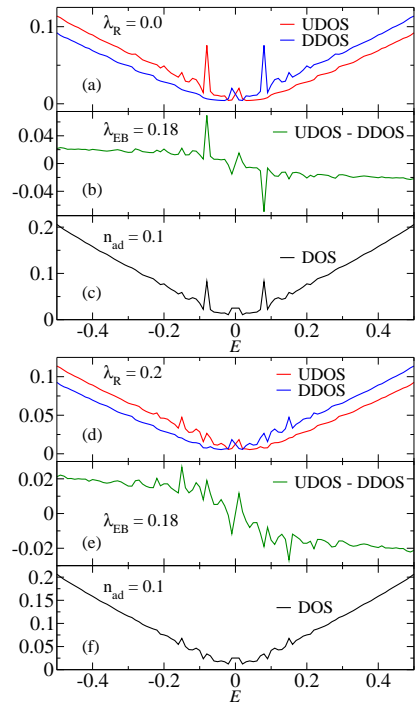

\begin{center}
\includegraphics[width=0.3\textwidth]{fig14a.eps}\\\includegraphics[width=0.3\textwidth]{fig14b.eps}
\caption{(Color online) The four terminal density of states (DOS) for the spin up and spin down electrons are plotted as a function of the Fermi energy (a) for $\lambda_R = 0$ and (d) for $\lambda_R = 0.2$. The difference between DOS$\uparrow$ and DOS$\downarrow$ are plotted as a function of $E$ (b) for $\lambda_R = 0$ and (e) for $\lambda_R = 0.2$. The behaviour of total DOS is plotted as a function of the Fermi energy (c) for $\lambda_R = 0$ and (e) for $\lambda_R = 0.2$. for clarity we consider only a single adatom concentration, $n_{ad} = 0.1$.}
\label{4t_dos}
\end{center}
\end{figure}

The DOS plots in the absence of Rashba SOI are shown in Fig.\ref{4t_dos}(a-c) and in the presence of Rashba SOI in Fig.\ref{4t_dos}(d-f). UDOS and DDOS are antisymmetric to each other as we have observed in case of two terminal case. the magnitude of UDOS and DDOS are higher at higher values of $E$ as shown in Fig.\ref{4t_dos}(a) and Fig.\ref{4t_dos}(d) and in both the cases, that is, in presence and absence of Rashba SOI, their magnitudes seem to be independent of the strength of Rashba SOI. The difference between UDOS and DDOS, DiffDOS (= UDOS - DDOS) is antisymmetric in nature as a function of the Fermi energy and has somewhat larger oscillatory behaviour in presence of RSOI than that in the absence of RSOI. Also the total DOS is symmetric about $E=0$. Basically all the three quantities are in good agreement with Eq.\ref{dos} both in the absence and presence of Rashba SOI.

\begin{figure}[h]
\begin{center}
\includegraphics[width=0.23\textwidth]{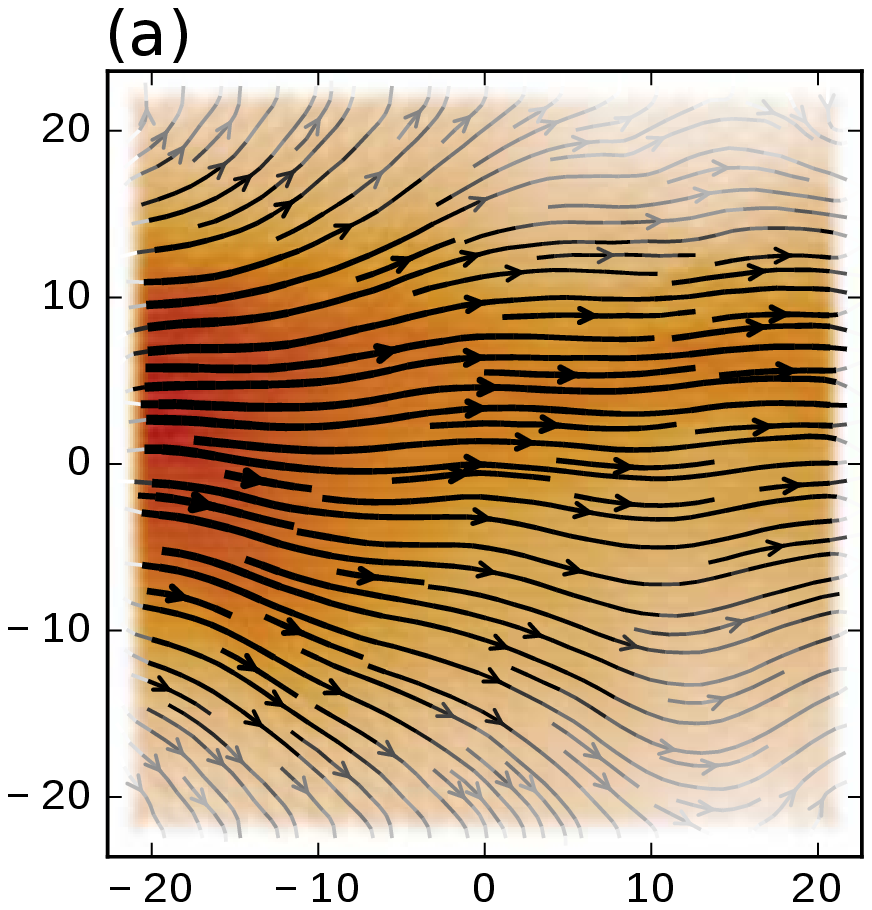}\includegraphics[width=0.23\textwidth]{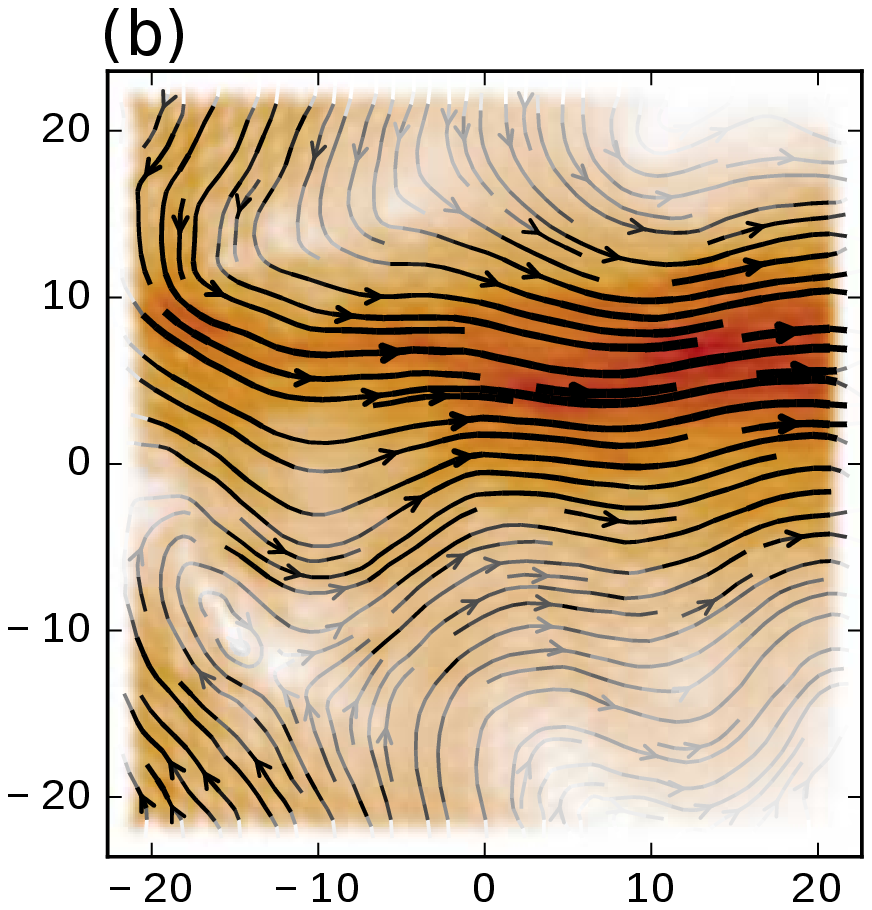}
\caption{(Color online) (a) For the four terminal case, the local charge current, $J_0$ and (b) the $z$-component of the local spin current are shown in the absence of Rashba SOI. Both the figures are obtained for the adatom concentration $n_{ad} = 0.1$ and we set the Fermi energy at $E = -0.09$.}
\label{4t_current}
\end{center}
\end{figure}

Finally, the local charge current and the $z$-component of the local spin current are shown in Fig.\ref{4t_current}(a) and Fig.\ref{4t_current}(a) respectively in the absence of Rashba SOI. For the local charge current, it is observed that as though the transverse leads are voltage probes, the charge current is flowing between lead 1 and lead 3. In another word, the charges are trying to accumulate at the transverse edges of the sample and as a result less charge current flows between terminals 1 and 2. This in turn reduces the charge conductance in case of a 4T device than the 2T case. The $z$-component of the local spin current is clearly flowing from terminals 1 to 3 and between terminal 1 to 4. As a result the non-zero $G_{SH}^z$ occurs due to presence of the exchange field.

\begin{figure}[h]
\begin{center}
\includegraphics[width=0.2\textwidth]{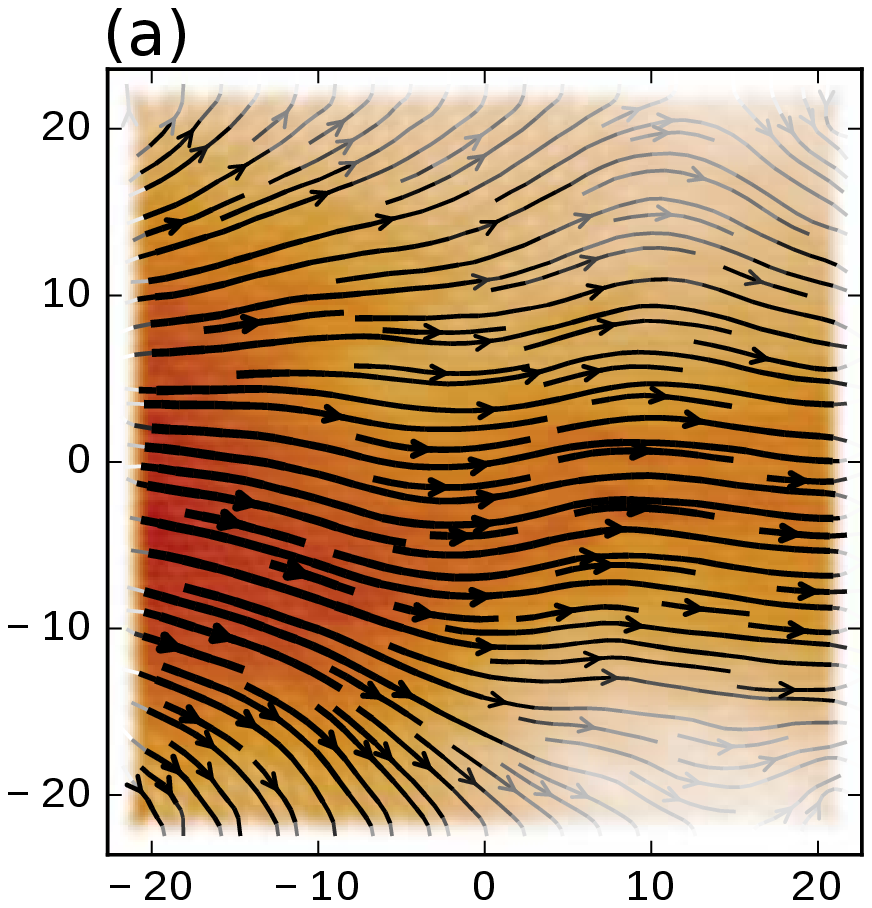}\quad\includegraphics[width=0.2\textwidth]{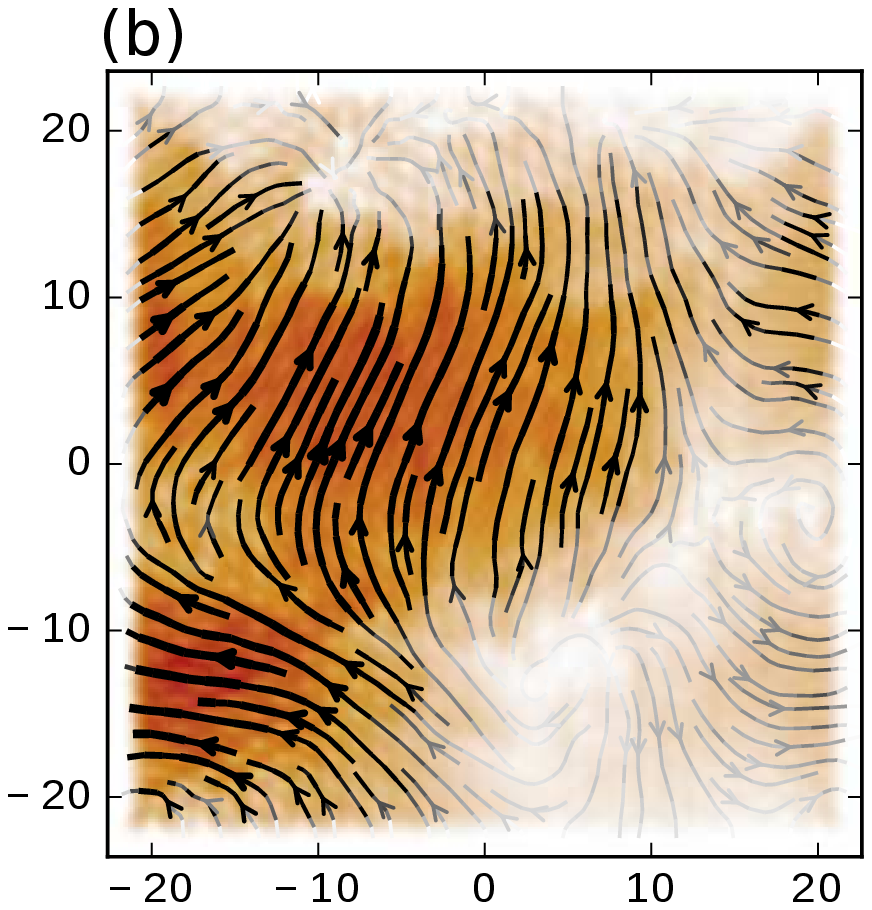}
\includegraphics[width=0.2\textwidth]{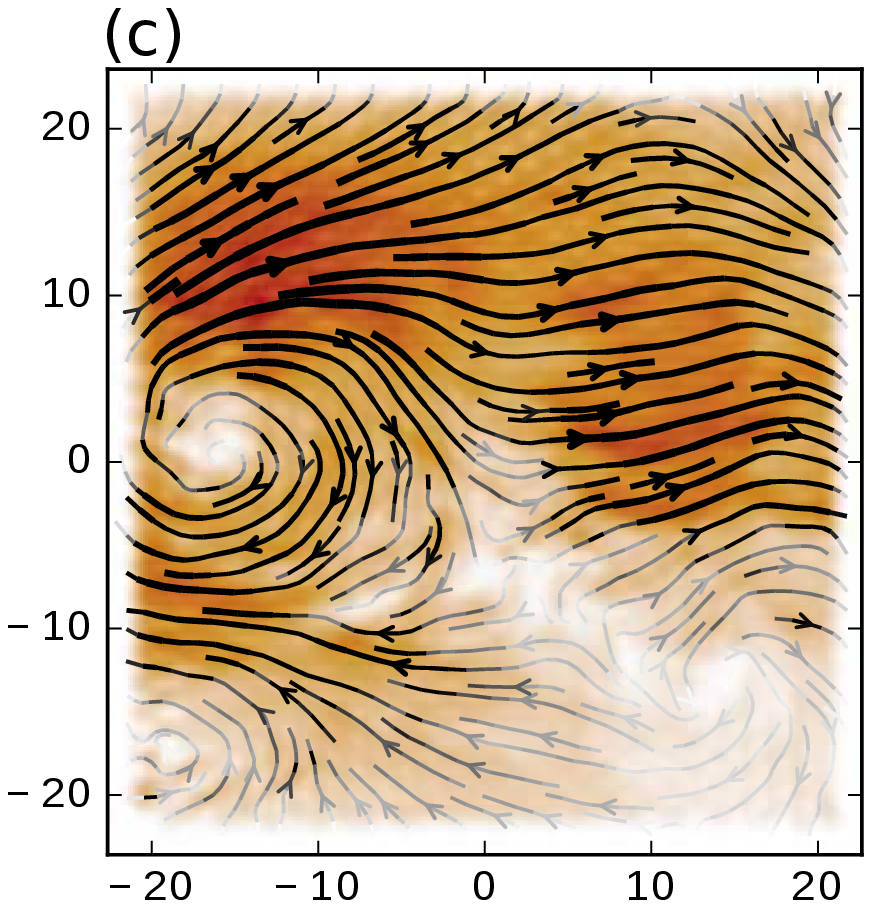}\quad\includegraphics[width=0.2\textwidth]{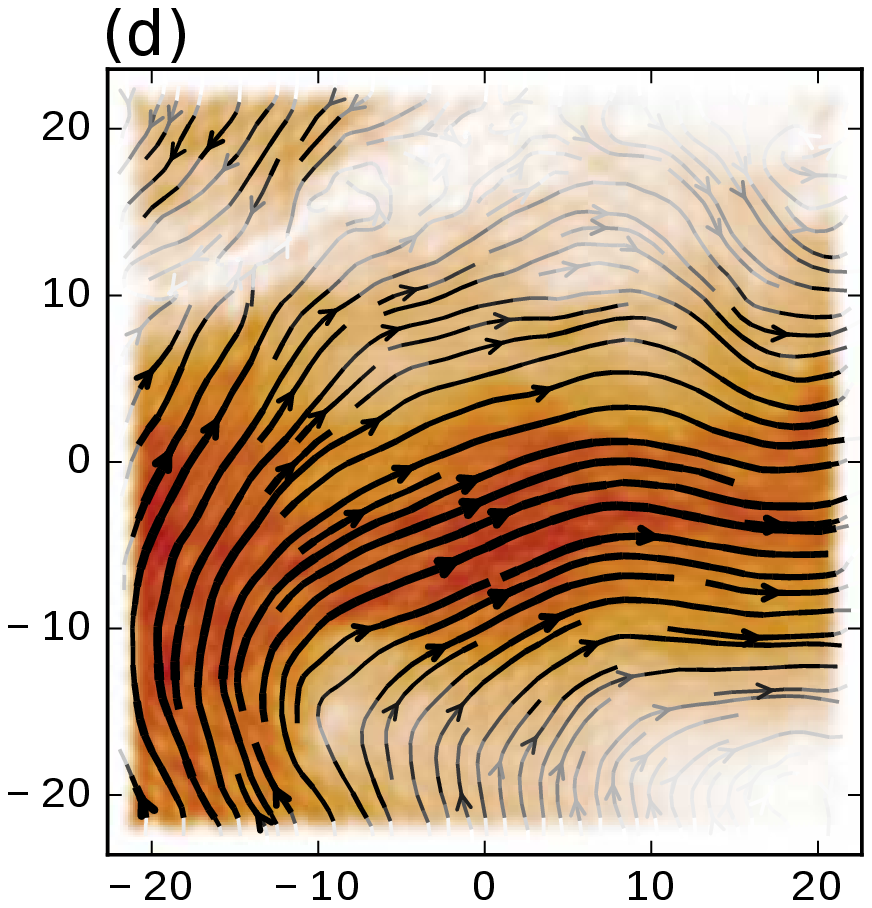}
\caption{(Color online) (a) The local charge current $J_0$, (b) the $x$-component of the local spin current, $J_x$, (c) the $y$-component of the local spin current, $J_y$ and (d) the $z$-component of the local spin current, $J_z$ are shown in the presence of Rashba SOI with strength $\lambda_R = 0.2$. The strength of the exchange field is $\lambda_{EB} = 0.18$. We set the Fermi energy at $E = -0.09$ and the adatom concentration in the present case is $n_{ad} = 0.1$.}
\label{4t_current_lr}
\end{center}
\end{figure}

The local charge and spin currents are shown in Fig.\ref{4t_current_lr} in presence of Rashba SOI. The local charge current, $J_0$ flows between the left and right leads (terminals 1 and 2 respectively) as shown in Fig.\ref{4t_current_lr}(a). We observe that $J_0$ is almost independent of the Rashba SOI. This explains why the charge conductance is independent of the Rashba SOI as shown in Fig.\ref{4t_cond}(a) and Fig.\ref{4t_cond}(b). 

Fig.\ref{4t_current_lr}(b), (c) and (d) show respectively local currents, namely $J_x$, $J_y$ and $J_z$. We observe that the magnitude of the $z$-component of the spin Hall conductance (see Fig.\ref{4t_gsz}(a-c)) is one order magnitude lower than any one of the components of SHC in presence of Rashba SOI, which can be explained as following. From Fig.\ref{4t_current_lr}(b-c), the number of paths between terminal 1 and either terminals 3 or 4 are more than the absence of Rashba SOI case as shown in Fig.\ref{4t_current}(b). As a result, the amount of spin current will be less in the absence of Rashba SOI.

\section{\label{sec6}Conclusion}
In the present work we have studied the behaviour of the charge and spin transport properties in graphene nanoribbon with magnetic adsorbates both in case of two terminal and a four terminal GNR. Specifically for the two terminal case, we study the charge and spin polarized conductance and for the four terminal case the spin Hall conductance. In all the cases we found that the charge conductance is symmetric about the zero of the Fermi energy, while the spin polarized conductance (for two terminal case) and spin Hall conductance (for four terminal case) are antisymmetric about the zero of the Fermi energy.The fluctuations of the charge and spin conductances show systematic behaviour, that is they increase with increasing adatom concentration. We also study the DOS behaviour and their behaviour are alike in both two and four terminal cases. The local charge current is found to be independent of the strength of Rashba SOI, while the three different components of the local spin currents are sensitive to Rashba SOI that is generated by the magnetic adatoms. Further the $z$-component of the spin polarized conductance for 2T GNR is larger by approximately a factor of 5 compared to its 4T counterpart, while the other two components are nearly same for the 2T and 4T devices. Of course a 4T device permits observations of a spin Hall conductance, which is absent in the case of a 2T GNR. Moreover the conductance properties of a 4T setup has lesser fluctuations.

The increase in the components of the spin polarized conductances in magnetic decorated GNRs with Rashba SOI signal a larger spin current flowing in the sample and hence  must have greater utility as possible spintronic devices.

\setcounter{secnumdepth}{0}
\section{ACKNOWLEDGMENTS}
SG gratefully acknowledges a research fellowship from MHRD, Govt. of India.
SB thanks SERB, India for financial support under the grant F. No: EMR/2015/001039.

\end{document}